# Performance Evaluation and Impact of Weighting Factors on an Energy and Delay aware Dynamic Source Routing Protocol

Jihen Drira Rekik[1], Leïla Baccouche[1] and Henda Ben Ghezala[1]

[1]RIADI-GDL laboratory, ENSI National school of computer sciences

Manouba University, 2010 Manouba, Tunisia

## Abstract

*Typical applications of the mobile ad-hoc network, MANET, are in disaster recovery operations which have to respect time constraint needs. Since MANET is affected by limited resources such as power constraints, it is a challenge to respect the deadline of a real-time data. This paper proposes the Energy and Delay aware based on Dynamic Source Routing protocol, ED-DSR. ED-DSR efficiently utilizes the network resources such as the intermediate mobile nodes energy and load. It ensures both timeliness and energy efficiency by avoiding low-power and overloaded intermediate mobile nodes. Through simulations, we compare our proposed routing protocol with the basic routing protocol Dynamic Source Routing, DSR. Weighting factors are introduced to improve the route selection. Simulation results, using the NS-2 simulator, show that the proposed protocol prolongs the network lifetime (up to 66%), increases the volume of packets delivered while meeting the data flows real-time constraints and shortens the end-to-end delay.*

## 1 Introduction

A Mobile Ad-hoc NETwork (MANET) has become increasingly popular due to its autonomic and infrastructure-less properties of dynamically self-organizing, self-configuring and self-adapting. With MANET, mobile nodes can move and access data randomly at anytime and anywhere. There is no need for fixed infrastructure. Mobile nodes such as PDA or laptops are connected by wireless links. They may act as a host and as a router in the network. They are characterized by their reduced memory, storage, power and computing capabilities. We classify the mobile nodes into two groups: small mobile hosts (SMH) which have a reduced memory, storage, power and computing capabilities and large mobile hosts (LMH) equipped with more storage, power, communication and computing facilities than the SMH.

MANET covers a large range of applications such as military operations where common wired infrastructures are not directly reachable to provide communication due to limited provision of this facility in those settlements. We focus especially on real-time applications where a number of them, including defence applications, have to respect time constraint in order to update wounded or positions of soldiers and enemies, get enemy map position or find medical assistance.

Real-time applications require their flows to be treated not only correctly but also within their deadlines. However, the workload of real-time applications is unpredictable which may lead the mobile nodes to become quickly overloaded. Since in MANET the mobile nodes are power limited and require energy for computing as well as routing the packets, the performance of a real-time application highly depends on the lifetime of mobile nodes. In fact, the energy depletion of mobile nodes may lead to interruptions in communications. Therefore, the real-time data may miss their deadlines. So respecting the deadline cannot be guaranteed neither with exhausted energy resources nor with overloaded intermediate mobile nodes. The main problem is to choose the QoS aware routing protocol to route real-time data with respect to their deadlines within MANET constraints.

Currently, most MANET research has focused on routing and connectivity issues [3] [7] in order to cope with the dynamism of such networks. Just solving the problem of connectivity is not sufficient for using MANET. Some routing protocols are extended to support the quality of service, QoS. To determine a route, a QoS routing protocol considers QoS requirements of the traffic data (such as maximum bandwith availability, minimum end-to-end delay and so on.) and resources availability (such as maximum residual energy, etc.), too.

Based on dynamic source routing (DSR) [5], we introduce the Energy and Delay aware Dynamic Source Routing protocol (ED-DSR) for MANET. ED-DSR is a routing protocol that considers the energy efficiency and load capacities in selecting route while focusing on the delay guarantee and the overall network performance. In ED-DSR, the route selection is done according to residual energy and queue load of intermediate nodes, too. ED-DSR allows the packets of real-time data to be routed before the expiration delay. Simulation results show that ED-DSR outperforms the basic routing protocol, DSR, in providing longer network lifetime and lower energy consumption per bit of information delivered. In addition, it minimizes the end-to-end delay and upgrades the rate of packets delivered.

The rest of the paper is organized as follows: in the second section, we present the basic routing protocol. Then, we expose the related works in quality of service routing protocols. In the fourth section, we describe the proposed Energy and Delay-aware Dynamic Source Routing (ED-DSR) protocol. Detailed analysis of the performance difference is performed in sections 5 and 6.

## 2   THE DYNAMIC SOURCE ROUTING PROTOCOL, DSR

With routing protocols in mobile ad-hoc network, the mobile nodes search for a route to connect to each other in order to share the data packets. The routing protocols can be categorized into two, namely, table driven proactive like OLSR (Optimized Link State Routing Protocol) [9] and on-demand-driven reactive source initiated protocols like DSR [5]. The focus in this work is on reactive routing suitable to be deployed in a network with high mobility of the nodes where routes are created dynamically as and when required [10].

The Dynamic Source Routing Protocol, DSR, is an on-demand routing protocol [5]. It discovers routes between two nodes only when required which reduces the number of control packets. DSR is based on three phases: the route request, the route reply and the route selection. In route request, the source mobile node discovers routes to the destination node. The route reply returns the discovered route from the destination to the source mobile node. In route selection, the source mobile node selects the shortest route among the discovered routes.

The route discovery is based on two messages i.e. route request packet (RREQ) and route reply packet (RREP). When a mobile node wishes to send a message to a specific destination, it broadcasts the RREQ packet in the network. The neighbour nodes in the broadcast range receive this RREQ message and add their own address and again rebroadcast it in the network.

International Journal of Computer Science & Information Technology (IJCSIT) Vol 3, No 4, August 2011

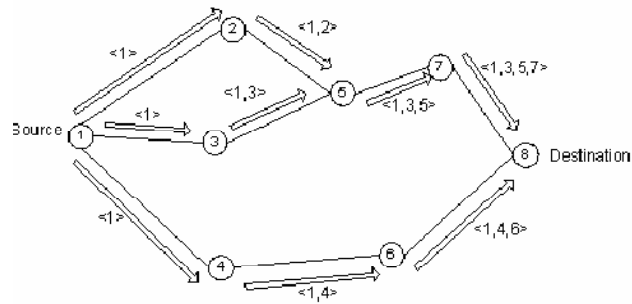

Figure 1 DSR RREQ packets

When the RREQ message reaches the destination, the route to the specific destination is yet defined. In fact, the message that reaches the destination has full information about the route. That node will send a RREP packet to the sender (source node) in order to have complete route information.

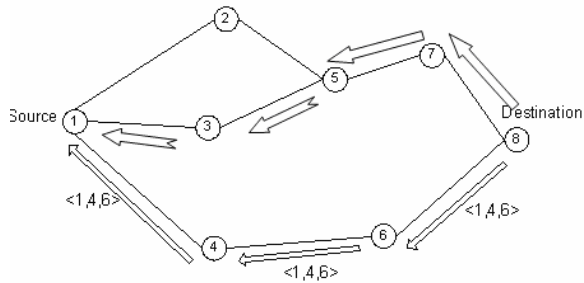

Figure 2 DSR RREP packets

The source node, among the discovered routes, selects the shortest one. The source node now has complete information about the route in its route cache and can start routing the data flows.

In DSR, the routes are stored in without any constraint on quality of services. The delay requirement is not considered to ensure that packets will reach their destinations before the deadlines. Furthermore, DSR doesn't contribute to reduce the power consumption of mobile node, alleviating the network partitioning problem caused by the energy exhaustion of these nodes.

## 3 RELATED WORK IN QUALITY OF SERVICE ROUTING PROTOCOLS

The requested quality of service, QoS, for a network depends on the characteristics of this network and the needs of its users. RFC 2386 [2] characterizes the QoS as a set of service requirements to be met by the network while transporting a flow from one source to a destination. These requirements can be expressed as a set of attributes pre-specified and measurable in terms of delay, jitter, bandwidth and packet loss. In MANETs, the QoS criteria should be adapted to the dynamic nature of network topology and limited battery resources.

The performance of the ad-hoc mobile network highly depends on the lifetime of mobile hosts. The network partition may lead to interruptions in communications, as in such conditions mobile nodes need to deliver their packets through intermediate nodes in order to intend destinations. Therefore, the lifetime of intermediate mobile nodes should be prolonged by conserving energy either at each node and for each connection request, too. Since most mobile hosts today are powered by batteries, efficient utilization of battery power assumes importance in MANET as the ad-hoc networks nodes are power limited and require energy for computing as well as routing the packets. Therefore, the early death of mobile nodes due to energy exhaustion may lead to the network partitioning and hence the disruption of service. In this environment, both the user and the data source will be moving, so finding a route from one mobile node to another mobile node is usual necessary before submitting a real-time data. Moreover applications in this environment are time-critical which require their flows to be executed not only correctly but also within their deadlines. A load balancing among mobile nodes should be provided while at the same time it will contribute to reduce the number of dropped packets (due to the deadline miss). Indeed, this technique ensures balance of energy consumption, too, among mobile nodes. In next subsections we give an overview of various proposed solutions. In the literature, a lot of QoS aware routing protocols have been proposed [6], [8] and [11]. To determine a route, QoS routing considers QoS requirements of the traffic flow and resources availability, too.

### 3.1 The Energy aware Multipath Routing Protocol, EMRP

The EMRP is an energy-aware multipath source routing protocol based on Dynamic Source Routing (DSR) [12]. It makes changes in the phases of Route Reply, Route Selection and Route Maintenance according to DSR. EMRP utilizes the energy and queuing information to select better routes. In *route reply*, each intermediate mobile node will stamp its current status in the RREP packet. Finally, the routing agent at the source node will collect the RREP. In *routes selection*, EMRP chooses the working set of routes from all available routes according to the following rules. First of all, EMRP calculates the cost of each available route according to the following equation:

$$W = \sum_{i=1}^{n} \left( \alpha \times W_{energy}^i + \beta \times W_{queue}^i \right). \tag{1}$$

Where $W$ is the cost of the route and $W_{energy}^i$, $W_{queue}^i$ are the costs of node $i$ considering the energy and queue length respectively. $\alpha$ and $\beta$ are the costing factors which normalize $W_{energy}^i$ and $W_{queue}^i$. A route is selected based on minimum values of $W$. $W_{energy}^i$ is calculated as follows:

$$W_{energy}^i = \left( \frac{P_{tx}^i}{E_{remain}^i} + \frac{P_{rx}^{i+1}}{E_{remain}^{i+1}} \right) + \left( 1 + N_{retrans}^i \right). \tag{2}$$

Where $P_{tx}^i$ and $P_{rx}^i$ are the transmitting energy cost from node $i$ to the next-hop node $i+1$ and the receiving energy cost of the next-hop node $i+1$, respectively. $W_{energy}^i$ is a function depending of the distance and remaining energy of node $i$ and the next-hop node $i+1$. More remaining energy and shorter distance indicate less $W_{energy}^i$. $W_{queue}^i$ is given below:

$$W_{queue}^i = log(1 + N_{queue}^i). \tag{3}$$

Where $N_{queue}^i$ is the queue length at node $i$. $W_{queue}^i$ depends on the queue length along the current route. If there are more packets in the queues along the route, the transmission will inevitably suffer a longer delay.

### 3.2 The Real-Time Dynamic Source Routing, RT-DSR

The RT-DSR is based on the expiration delay to deadline [13]. It makes changes in the phases of route request and route reply. In *route request*, a route request RREQ packet is broadcasted with the expiration delay to deadline.

The route request packet is accepted for a new flow only if the new packet can reach the destination before the expiration delay (4).

$$E_k^{i-1} - T_{TL} - T_{TS} > 0. \tag{4}$$

Where $E_k^{i-1}$ is the remaining time of the expiration delay to deadline, for the traffic k, received from the node *(i-1)*. $T_{TL}$ is the local processing time of any message; $T_{TS}$ is the transmission time between two neighboring nodes in the worst case remaining times. The delay of each real-time data in the queue, already admitted, shouldn't be altered by the newest one.

$$\forall j, 1 \leq j \leq res, E_j^{t-1} - T_{TL} - T_{TS} > 0. \tag{5}$$

Where *res* is the number of real-time data already admitted in the node.

In *route reply*, the second phase allows reserving resources of the discovered route, by the first phase. Each intermediate node reserves the resources, saves the remaining time value to deadline and sends the confirmation packet to the next node until reaching the source node.

### 3.3 The Adaptive Link Weight Routing Protocol, ALW

In [1], the ALW is a routing protocol based on three quality of service parameters: $K_1$ bandwidth (data rate), $K_2$ link delay (latency) and $K_3$ route lifetime (minimum battery lifetime of nodes in the route). These QoS parameters are defined as weighting factors according to the application requirements. They are integrated into the cost function used in route selection phase. The link weight cost function is calculated using the following equation for selecting a route:

$$Link\ weight = (K_1 \times Bandwith) + (K_2 \times Delay) + (K_3 \times Node_{Lifetime}).$$

Where $K_1 + K_2 + K_3 = 1$.

Different types of application having dissimilar QoS requirements are defined in [1]. The weighting factors are defined according to the application requirements. The route selection process is adaptive and closely matches the application requirements.

Table 1. Defined weighting factor according to the application requirements

| Applications | $K_1$ | $K_2$ | $K_3$ |
|---|---|---|---|
| Video conference | 0,5 | 0,4 | 0,1 |
| FTP | 0,5 | 0,3 | 0,2 |
| Messaging service | 0,1 | 0,4 | 0,5 |
| Default (Optimum) | 0,33 | 0,33 | 0,33 |

Different routes may be selected between the same source and destination nodes relative to the application requirements where different types of applications are hosted at these nodes.

### 3.4 Limits of QoS routing protocols

The EMRP solution proposes multipath routing protocol. It provides routes that reduce the intermediate mobile nodes power consumption, alleviating the network partitioning problem caused by the energy exhaustion of these nodes in order to assure successful data packet transmission. However, the exhaustible energy battery is not the only indicator for route selection and a power control scheme. The number of packets in each node's queue, along the route, doesn't reflect the local processing time. In fact, each packet has its proper execution time which varies. Thus, the packet handing will inevitably suffer a longer delay and therefore the energy exhaustion of these nodes; while there are other nodes with less energy but where their queues require less time to be treated. The route selection should be done according to energy and more queuing information, in terms of queue length and local processing time of each previous flow, too.

The RT-DSR purpose is to reserve resources in order to meet the deadlines but the proposed routing protocol must also take care that the resources are exhaustible. Indeed, choosing the same route to transfer all packets of real-time data through the reserved route may exhaust the energy of these nodes leading to the network partitioning problem. Moreover, the route selection criteria should consider that in the mobile ad-hoc network there are other traffics generated and they could take some joint nodes. The rules, under which packets are assigned to route, should improve the system performance in terms of real-time constraint and energy efficiency too.

## 4 THE ENERGY AND DELAY AWARE DYNAMIC SOURCE ROUTING: PROPOSED PROTOCOL

The ED-DSR proposed routing protocol considers the packet deadline (real-time constraint), the remaining energy of the forwarding nodes and the load at intermediate mobile nodes to deliver real-time data. Therefore, each packet should be transmitted from the source to the destination within the deadline. The basic working of our proposed protocol is as follows.

Each mobile node, before starting the transmission of real-time data, selects a suitable route between the source and the destination. The selected route should satisfy delay requirements, preserve energy consumption and avoid overloaded intermediate mobile nodes. Energy delay-dynamic source routing, ED-DSR, protocol is based on DSR. DSR is an on demand protocol ensuring the freshness of constructed route which is more suitable for the real-time data. Therefore, we opt to DSR as based protocol in our work. DSR discovers a route between two nodes, only when required which reduces the number of packets control. DSR is simple and flexible [4] which facilitates the implementation of our extension. Also, a route response packet sent back to the source can be used to incorporate real-time and energy constraints. The choice of the suitable route to transfer the real time data in ED-DSR is conditioned by three factors: the residual energy of nodes belonging to the route, the delay requirements of the real-time data and the load of the node's queue.

### 4.1 The Packet Format and The Process Chart

The ED-DSR uses two control packets: the Route Request packet: RREQ and the Route Reply packet: RREP. The format of our RREP packet is different from the original one used with DSR in order to introduce the QoS parameters for route selection. We modified the RREP packet format and added two extra fields in the packet format of DSR to store the route cost function, $C$, relative to the actual status of each intermediate mobile nodes and the delay cost function, $C_{Delay}$. The $C$ and $C_{Delay}$ are described in the next route selection subsection.

| Next Header | Reserved | Payload Length | |
|---|---|---|---|
| Options | Data | C | $C_{Delay}$ |

Figure 3 Modified RREP packet format of ED-DSR

Figure 4 represents the flow chart of ED-DSR highlighting ours contributions within bold chart.

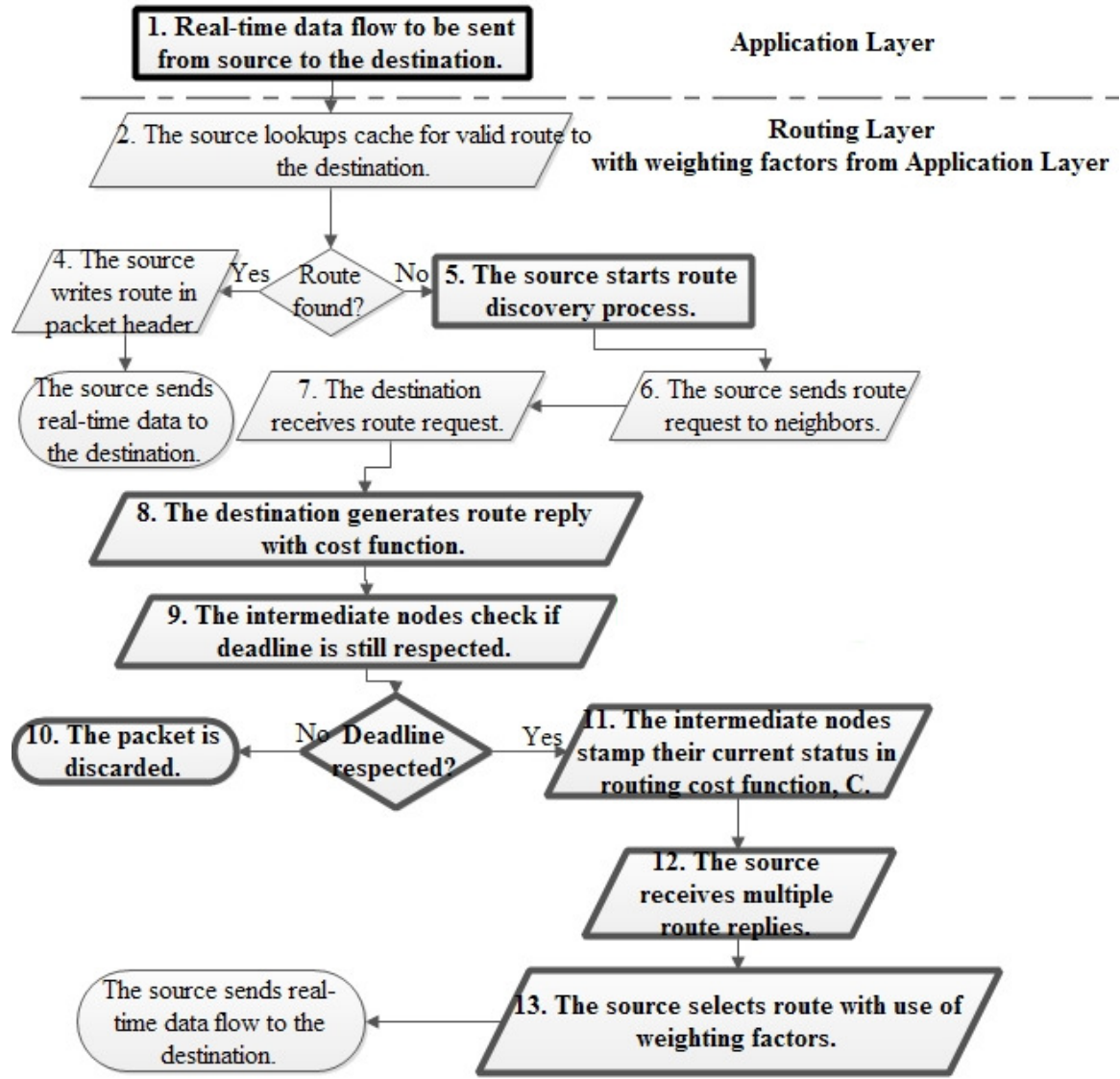

Figure 4 Flow chart of ED-DSR highlighting the route selection contribution

At each intermediate node i, the expiration delay is compared to the cost delay contained in the received packet RREP (Figure4, Process9). If the new link cost delay is expired, the packet is discarded. Otherwise, the packet is forwarded to the next intermediate node $i + 1$.

## 4.2  Route request

When transmitting a new real-time data, the source node checks its route cache first to see whether there are available routes to the destination node (Figure4, Process2). If routes are available, the protocol selects the suitable route according to the rules, which will be presented in next sub-section. Otherwise, the source node goes into the route discovery phase to request available route by broadcasting a RREQ packet to all other nodes (Figure4, Process6).

## 4.3  Route reply

When a destination node receives an RREQ, it returns back a Route REPly (RREP) packet to the source node (Figure4, Process8). Different from DSR, in ED-DSR, while an RREP packet is being sent back to the source node, each intermediate mobile node will stamp its current status in the RREP packet (Figure4, Process11). Finally, at the source node, the routing agent collects the RREP (Figure4, Process12). This status information is shown in Table 1, in which $i$ is the index for the mobile nodes.

Table 2 Information fields of RREP packets

| Information fields | Contents |
| --- | --- |
| $d_i$ | Distance to this node provided by the physical layer |
| $L_{queue}^i$ | Current length of queue, provided by the network layer. |
| $E_{remain}^i$ | Current remaining energy of this node, provided by the physical layer. |

ED-DSR calculates the cost of each available route according to the following equation:

$$C = \sum_{i=1}^{M}(\alpha \times C_{energy}^i + \beta \times C_{queue}^i + \gamma \times C_{delay}^i). \quad (6)$$

where $C$ is the cost of the route and $C_{energy}^i$, $C_{queue}^i$ are the costs of node $i$ considering the energy and queue length respectively. $\alpha$, $\beta$ and $\gamma$ are the factors which normalize $C_{energy}^i$, $C_{queue}^i$ and $C_{delay}^i$. $C_{energy}^i$ is calculated as follows:

$$C_{energy}^i = \left(\frac{d_i}{E_{remain}^i}\right). \quad (7)$$

$C_{energy}^i$ is a function depending of the distance and remaining energy of node $i$. More remaining energy and shorter distance indicate less $C_{energy}^i$. $C_{queue}^i$ is given below:

$$C_{queue}^i = log(1 + L_{queue}^i). \quad (8)$$

where $L_{queue}^i$ is the queue length at node $i$. $C_{queue}^i$ equation is calculated in the same manner as [12]. It is relative to the queue length along the current route.

If there are more packets in the queues along the route, the transmission will inevitably suffer a longer delay. $C_{queue}^i$ increases rapidly with $L_{queue}^i$.

$$C_{delay}^i = L_{queue}^i \times T_L^i + T_T^i \times N_{hops}. \quad (9)$$

where $L_{queue}^i$ is the queue length at node $i$, $T_L$ is the local processing time of any message in node $i$; $T_T$ is the transmission time between two neighboring nodes in the worst case remaining times and $N_{hops}$ is the number of hops.

$C_{delay}^i$ depends on the queue length and the local processing time of each packet along the current route.

Each packet should verify if it can reach the destination before the expiration delay (10). Otherwise, the node discards the route.

$$D_k > \sum_{i=1}^{N} C_{delay}^i. \qquad (10)$$

Where $D_k$ is the worst case execution time for the packet k.

### 4.4 Route selection

In ED-DSR, the source node waits a certain period of time to collect RREP packets from the destination node along various routes, which is exactly what DSR does. But, different from DSR, among selected routes, the source node selects one based on minimum value of $C$ (Figure4, Process13).

We introduce three weighting factors $w_{energy}^i$, $w_{queue}^i$ and $w_{delay}^i$ to each cost functions $C_{energy}^i$, $C_{queue}^i$ and $C_{delay}^i$, respectively, in order to improve the route selection. The route selection depends to the application requirements and is not fixed as DSR which always selects a route with minimum hops to the destination. The weighting factors reflect the requested QoS features such as remaining energy, queue load of intermediate mobile nodes or the delay constraint of data packets.

The cost function will be calculated as follows:

$$C = \sum_{i=1}^{M} \left( w_{energy}^i \times (\alpha \times C_{energy}^i) + w_{queue}^i \times (\beta \times C_{queue}^i) + w_{delay}^i \times (\gamma \times C_{delay}^i) \right).$$

Where $w_{energy}^i + w_{queue}^i + w_{delay}^i = 1$. $w_{energy}^i$ allows to privilege the intermediate mobile node with higher remaining energy and shorter distance, $w_{queue}^i$ allows to privilege the intermediate mobile node with less packets in the queue and $w_{delay}^i$ allows to privilege the intermediate mobile node with less delay cost function.

The cost function is calculated from the current status information of the intermediate mobile node. The route selection criteria are relative to the application requirements specified by the weighting factors.

## 5 THE SIMULATION MODEL

We have used the Network Simulator, NS-2 in our simulations. NS-2 is an object-oriented, event driven simulator. It is suitable for designing new protocols, comparing different protocols and traffic evaluations.

### 5.1 Simulation environment

We simulated a MANET with 10 - 100 nodes in a 1500m×500m. With a rectangle area, longer distances between the nodes are possible than in a quadratic area, i.e. packets are sent over more hops. Each node is equipped with an IEEE 802.11 wireless interface in a priority queue of size 50 that drops packets at the queue end in case of overflow. A traffic load between pair of source-destination (SMH-LMH) is generated by varying the number of packets per second on the constant bit rate - CBR. Each packet is 512bytes in size.

We defined two groups of mobile nodes according to their resource capacity SMH and LMH. At the beginning of simulation, SMH nodes start with a starting energy of 50 joules and LMH with 100 joules. Since we do not address the problem of consumed energy in idle state, we have only considered energy consumed in transmission and reception modes. As values, we have utilized 1.4 W for transmission mode and 1 W for reception mode. The mobile nodes move around the simulation area based on the random waypoint (RWP) mobility model, with a maximum speed of 2 m/s and a pause time of 10 seconds for SMH, which model a soldier mobility pattern and speeds of up to 20 m/s for LMH, which corresponds more to vehicular movements.

All results reported here are the averages for at least 5 simulation runs. Each simulation runs for 1000 s. During each run, we assume that the SMH mobile node 0 wants to send real-time data to LMH the last node with an expiration delay equals to 15 seconds for firm real-time data constraint and 25 seconds for higher expiration delay for soft real-time data constraint. Then, we observe the behaviour of the mobile nodes.

## 5.2    Performance criteria

Five important performance metrics are evaluated. They are used to compare the performance of the routing protocols in the simulation:

- **Real-time packet delivery in time ratio**: the ratio of the real-time data packet that are delivered in time to the destination to those generated by CBR sources.
- **Real-time packet delivery ratio**: the ratio of the real-time data packets delivered to the destination to those generated by CBR sources.
- **Mean end-to-end delay**: the mean end-to-end delay is the time between the generation of a packet by the source up to data packets delivered to destination.
- **Network lifetime** The network lifetime corresponds to the first time when a node has depleted its battery power.
- **Energy consumption per bit delivery** is obtained by dividing the sum of the energy consumption of the network by the number of successfully delivered bits.

## 6    RESULTS AND DISCUSSIONS

Several simulations are performed using the NS-2 network simulator. The NS-2 generates a trace files analyzed using a statistical tool developed in AWK. The performance study concerns the basic routing protocol DSR which refers to the classic DSR protocol and the proposed routing protocol ED-DSR which refers to our QoS routing protocol for two expiration delays 15s and 25s, which reflect respectively firm and soft real-time deadline constraint.

## 6.1    Network performance

We propose here to study the impact of traffic load between pair of source-destination (SMH-LMH) by varying the number of packets per second on the CBR streams. The following figures show performance evaluation of DSR and ED-DSR protocols related to {5, 9, 10, 12, 15, 20} packets per second on the CBR streams for 50 mobile nodes.

We evaluate three metrics, namely, the rate of real-time packets that are delivered in-time (where the deadline constraint is still respected), the rate of real-time packets delivered and the average of end-to-end delay.

International Journal of Computer Science & Information Technology (IJCSIT) Vol 3, No 4, August 2011

### 6.1.1 Real-time packet delivery

Firstly, we observe and compare the variation of the ratio of all delivered packets regardless of compliance with the real-time constraints and the ratio of delivered packets in-time, which respects the real-time constraint, while the data rate of the CBR flow is increased.

In figure 5, we observe that DSR provides good performances; however, DSR's packet delivery ratio includes all packets that have reached the source node and where the deadline is not guaranteed for all packets received, as shown in figure 6. The ED-DSR packet delivery ratio reflects the packets that have respected their real-time deadline constraint and will be handled in time.

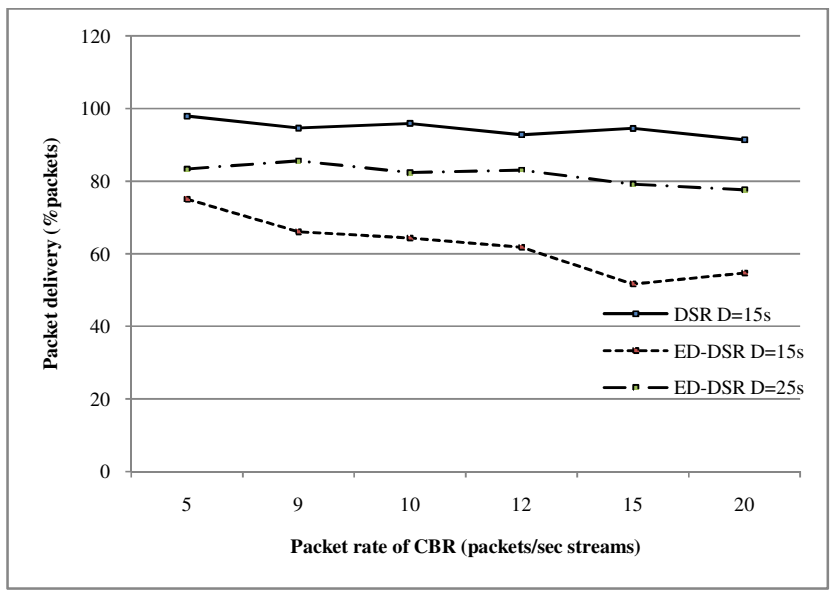

Figure 5. Real-time packets delivery ratio

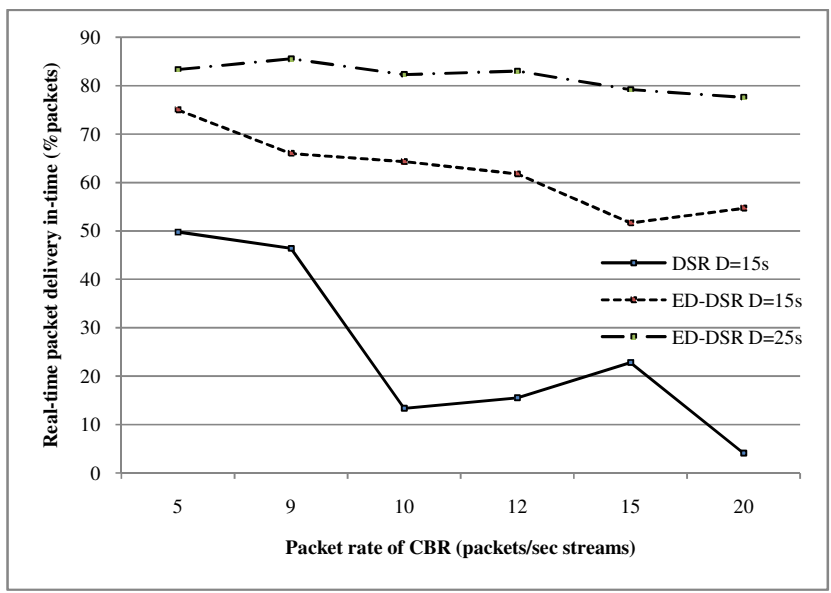

Figure 6. Real-time packets delivery in time ratio

In both figures 5 and 6, with ED-DSR, the ratio of real-time packet delivery at source node is the same. In fact, with ED-DSR, the real-time packets that expire their deadlines are discarded by the intermediate mobile nodes. Each intermediate mobile node verifies if the route response packet RREP respects or not the real-time constraint before reaching the source node. Otherwise, the RREP packet will be discarded. Therefore, the MANET will avoid the network overloading with packets that have expired their deadline in order to reduce energy consumption and alleviate network load. However, with DSR, the real-time constraint is not guaranteed especially as the packet rate value increases.

With firm real-time constraint, where D=15s, we note that packet delivery ratio in time decreases but stills stationary and better than DSR. The ratio of the packets sent within the compliance of its real-time constraint is over 50%. ED-DSR offers best performance for delivering real-time packets in time with soft real-time constraint, where D=25s.

### 6.1.2 End-to-end delay guarantee

Another commonly used metric is the average end-to-end delay. It is used to evaluate the network performance. As shown in figure 7, for low traffic (approximate to 5packets/sec), the packet end-to-end delay results experienced by both protocols are close. It implies that the delay is respected when the communication load is low. When the communication load increases, a number of packets are dropped, the route discovery is restarted and the packet delay increases with DSR. It indicates that packet delay is sensitive to the communication load and is basically dominated by the queue delay. However, with ED-DSR, the average end-to-end delay stills low. The network overloading is avoided by discarding the packets that expire their deadline and thus alleviates the load of mobile node queue.

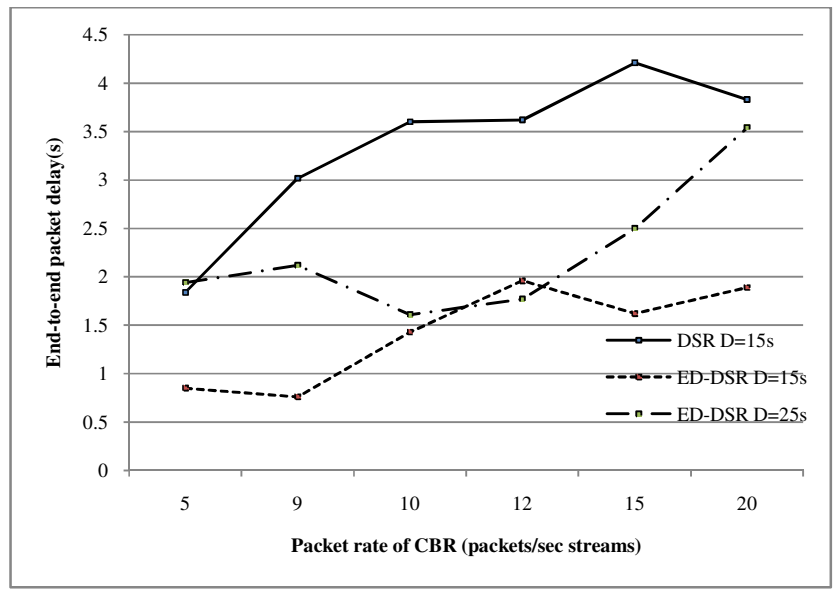

Figure 7. End-to-end packet delay

Our proposed protocol selects different routes depending on the cost function, thereby avoiding overloaded intermediate mobile nodes in the network and thus reducing the delay for packets. For high network traffic (up to 9packets/sec), our protocol gives much improved performance.

International Journal of Computer Science & Information Technology (IJCSIT) Vol 3, No 4, August 2011

We also, note that the average end-to-end delay is better for lower deadline (D=15s, firm real-time deadline constraint). In fact, ED-DSR selects route which reduces the transmission delay in order to respect the deadline.

## 6.2 Energy efficiency

In this section, we focus especially on the impact of our proposed routing protocol ED-DSR on the energy efficiency guarantees. Two metrics: the network lifetime and the energy dissipation are used. We study the impact of traffic load between pair of source-destination (SMH-LMH) by varying the number of packets per second on the CBR connection for 50 mobile nodes. Then, we study the impact of network density on the mobile ad-hoc routing protocols performance. This criterion is simulated by varying the number of the mobile nodes between 10 and 100 {10, 20, 30, 50, 70, 100} with 10 packets per second on the CBR streams. We focus especially on the impact of our proposed protocol ED-DSR on energy efficiency guarantees.

### 6.2.1 Network lifetime

Firstly, we observe the variation of the network lifetime while the data rate of the CBR flow is increased. Figure 8 shows the simulation results on small mobile host lifetime comparing ED-DSR and DSR under various traffic loads, while the data rate of the CBR flow are increased.

We can see that networks running ED-DSR live longer than those running DSR, especially for high network traffic (up to 9packets/sec).

As evident by the graph, our ED-DSR is little bit as efficient as DSR with low connection rate and much better in high traffic load.

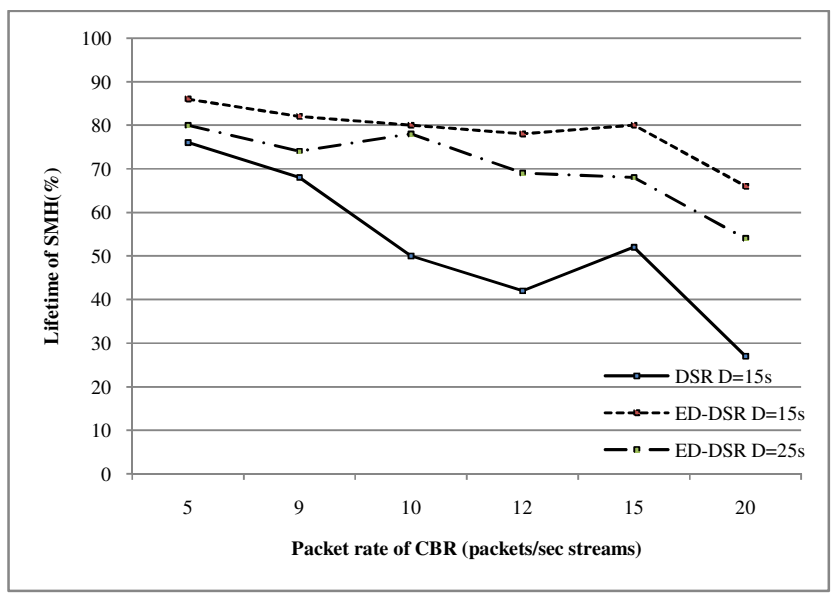

Figure 8. The network lifetime of SMH nodes for different traffic loads

By avoiding the network overloading with packets that have expired their deadlines and selecting routes that minimize energy cost, ED-DSR alleviates network load and reduces energy consumption, too.

International Journal of Computer Science & Information Technology (IJCSIT) Vol 3, No 4, August 2011

DSR network lifetime is low in approximately all cases in comparison to ED-DSR since DSR generates typically more routing overhead than ED-DSR.

Next, we observe the variation of network lifetime while the number of nodes is increased. Figure 9 shows the simulation results on SMH lifetime comparing ED-DSR and DSR.

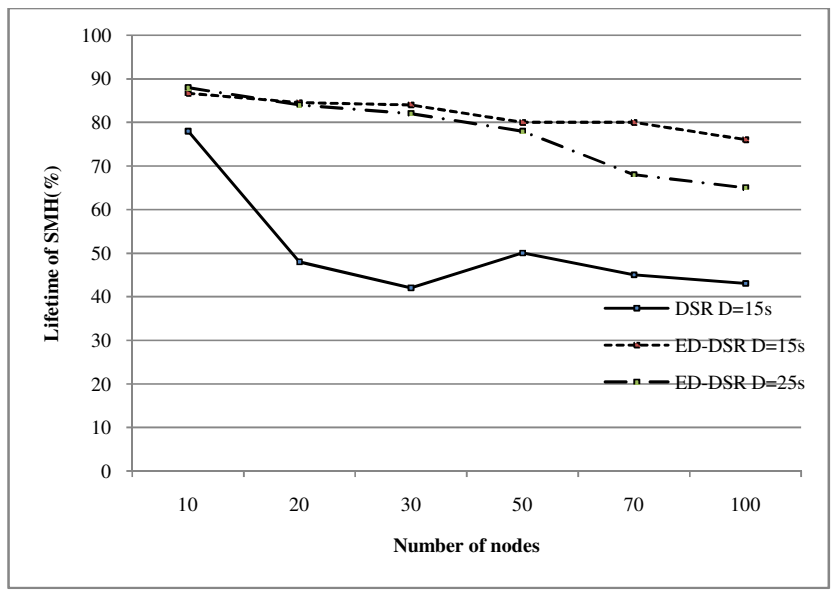

Figure 9. The network lifetime of SMH nodes for different node densities

Small mobile host, SMH, lifetime diminution according to node density augmentation is justified by the increase of generated routing overhead. Although the generated routing overhead has also increased in DSR, but this do not lead to an augmentation of its network lifetime. Nevertheless, DSR network lifetime is low in approximately all cases in comparison to ED-DSR since DSR generates typically more routing overhead than ED-DSR. In fact, in route selection, our proposal algorithm utilizes the network resources in terms of node energy and node load in order to balance traffic load. It ensures energy efficiency, up to 66%, by avoiding low-power node and busy node.

### 6.2.2 Energy dissipation

Figure 10 demonstrates the average energy consumption per bit delivery reflecting the global energy consumption in the network.

We see that ED-DSR outperforms DSR under different traffic loads, which is mainly due to the benefit of power control in the MAC layer. The excess packets inevitably introduce more collisions to the network, wasting more energy. ED-DSR chooses alternative routes, avoiding the heavily burdened nodes, thus alleviating the explosion in average energy consumption.

ED-DSR average energy consumption is lower than DSR average energy consumption under all packet rate conditions (over 9packets/sec) because ED-DSR selects path that minimize cost function.

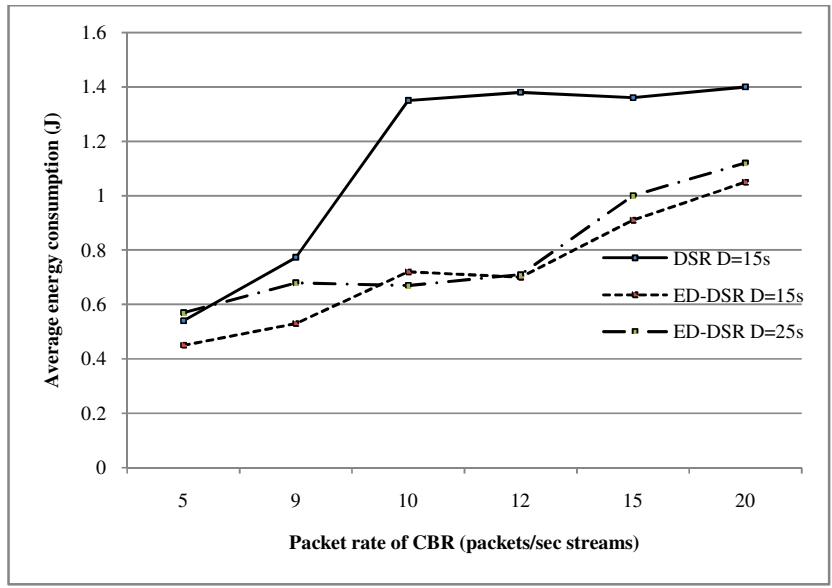

Figure 10. The average energy consumption per bit for different traffic load

Changing expiration delay for different packet rate has not a significant impact on average energy consumption of ED-DSR.

Figure 11 gives an idea about the global average energy consumption in the network for both protocols DSR and ED-DSR with different network densities.

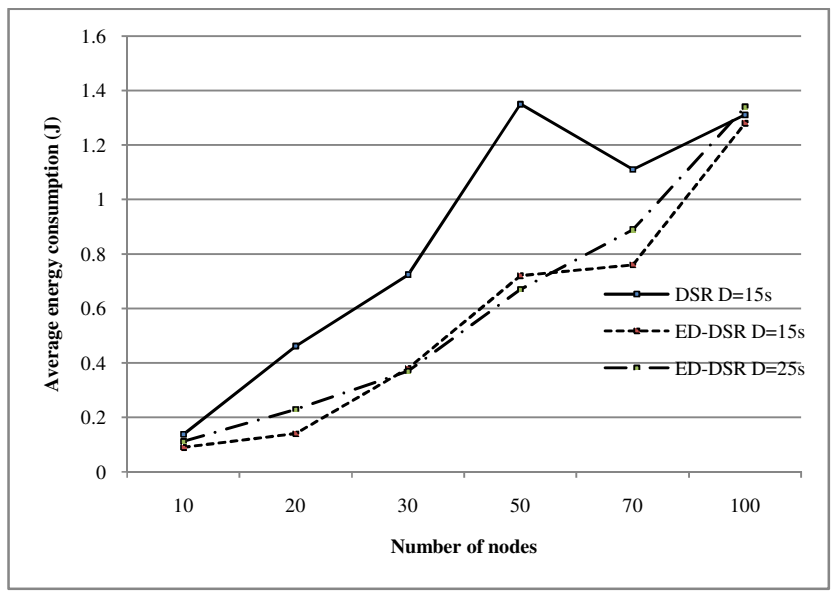

Figure 11. Average energy consumption per bit for different node densities

Increasing node density leads to an augmentation of collisions risk (consequently to more retransmission attempts) and to a growth in number of exchanged control packets. All those factors cause more energy dissipation for both protocols. ED-DSR average energy consumption is lower than DSR average energy consumption under all density conditions because ED-DSR

239

selects path that minimize cost function. Thus, its global energy consumption remains lower than DSR one. Changing expiration delay for different node densities has not a significant impact on average energy consumption of ED-DSR.

### 6.3 Impact of weighting factors

In this section, we propose to study the impact of weighting factors on the proposed routing protocol ED-DSR performance by varying the $w^i_{energy}$, $w^i_{queue}$ and $w^i_{delay}$ defined at the application layer. In table 3, the first line ED-DSR$_1$ privileges the energy cost function; the second one privileges the delay cost function for delay sensitive application and the last line is the default choice which the three weighting factors have similar opportunity in route selection phase.

Table 3 Weighting factors

|  | $w^i_{energy}$ | $w^i_{queue}$ | $w^i_{delay}$ |
|---|---|---|---|
| ED-DSR$_1$(Energy aware) | 0,6 | 0,2 | 0,2 |
| ED-DSR$_2$(Delay aware) | 0,2 | 0,2 | 0,6 |
| ED-DSR$_3$(The default) | 0,33 | 0,33 | 0,33 |

The following figures show the performance evaluation of ED-DSR protocol related to {5, 9, 10, 12, 15, 20} packets per second on the CBR streams for 20 mobile nodes.

We evaluate three metrics, namely, the rate of real-time packets that are delivered in-time (where the deadline constraint is still respected), the average of end-to-end delay and the average of energy consumption. The ED-DSR is monitored especially for firm real-time deadline constraint where the expiration delay is 15s.

#### 6.3.1 Packet delivery

The ED-DSR packet delivery ratio reflects the packets that have respected their real-time deadline constraint and will be handled in time. In figure 12, we note that the default ED-DSR (ED-DSR$_3$) gives approximately close results as the delay-aware routing protocol (ED-DSR$_2$).

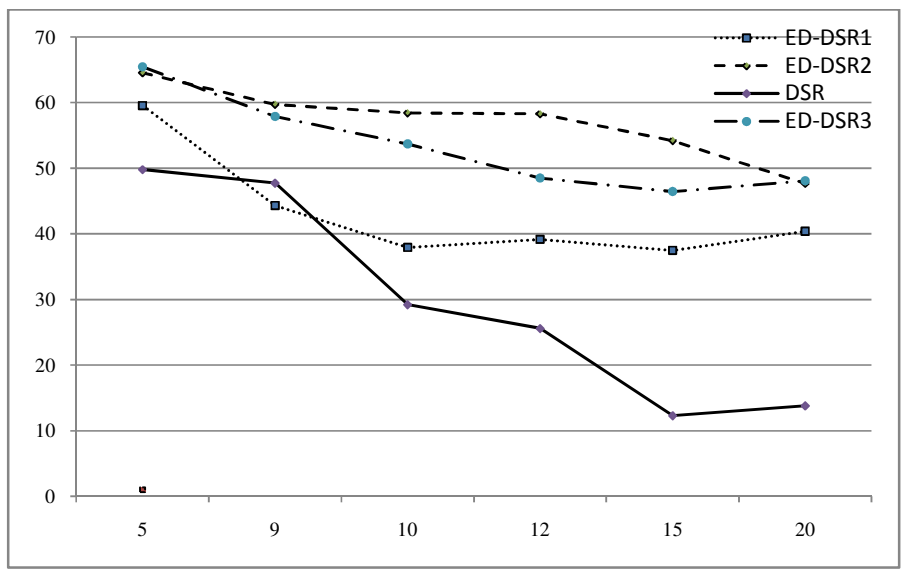

Figure 12. Real-time packets delivery in time ratio (20 mobile nodes)

International Journal of Computer Science & Information Technology (IJCSIT) Vol 3, No 4, August 2011

With ED-DSR$_1$ (the energy aware routing protocol), the ratio of the packets sent within the compliance of its real-time constraint is stable and is around 40 for high traffic load (over 9 packets/s). However, with DSR, the real-time constraint is not guaranteed especially as the packet rate value increases.

### 6.3.2  End to end delay guarantee

As shown in figure 13, we note that for all routing protocols the delay is respected when the communication load is low. When the communication load increases, a number of packets are dropped, the route discovery is restarted and the packet delay increases. It indicates that packet delay is sensitive to the communication load.

For high network traffic (up to 9packets/sec), the proposed protocol ED-DSR gives more improved performance than DSR.

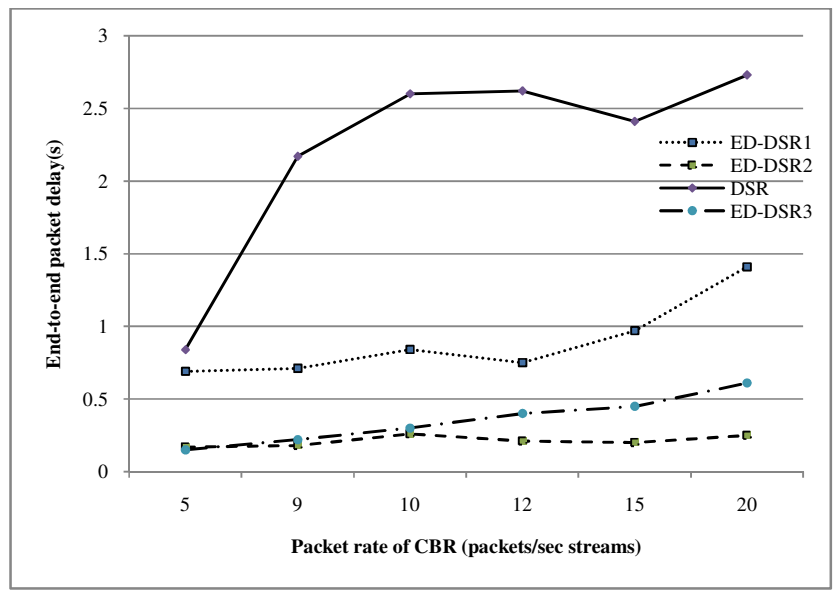

Figure 13. End-to-end packet delay (20 mobile nods)

The packet end-to-end delay results experienced by the delay aware (ED-DSR$_2$) and the default (ED-DSR$_3$) routing protocols are close and still low comparing to the energy aware (ED-DSR$_1$) and DSR routing protocols. In fact, the network overloading is avoided by discarding the packets that miss their deadline and thus reduces the load of mobile node queue. ED-DSR$_2$ privileges and selects route which reduces the transmission delay in order to respect the deadline.

### 6.3.3  Average energy consumption

Figure 14 demonstrates the average energy consumption per bit delivery. It gives an idea about the global energy consumption in the network comparing ED-DSR with different weighting factors and DSR under various traffic loads.

We see that ED-DSR highly outperforms DSR under different traffic loads, which is mainly due to the benefit of power control in the MAC layer, as proved in subsection 6.2.2.

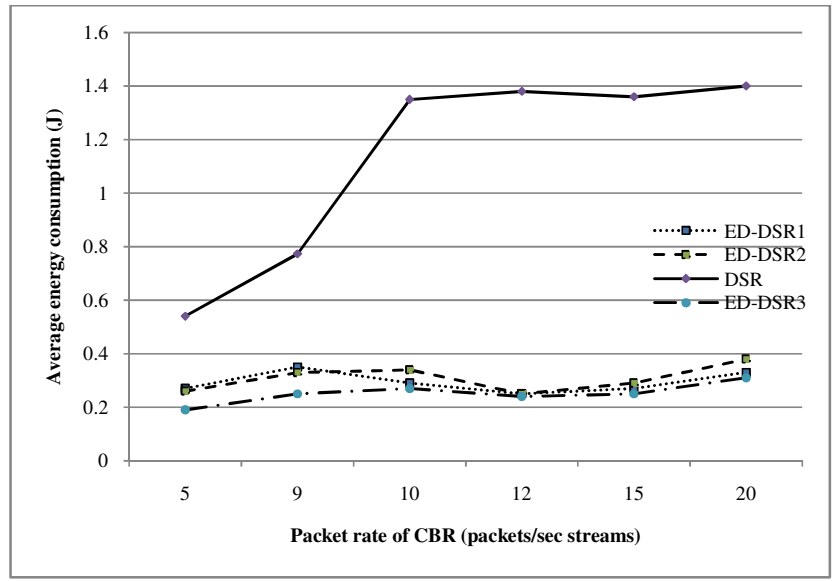

Figure 14. The average energy consumption per bit for different traffic load (20 mobile nodes)

Changing weighting factors (ED-DSR$_1$, ED-DSR$_2$ or ED-DSR$_3$) for different packet rate have approximately no significant impact on average energy consumption of ED-DSR, especially for high traffic load (up to 9packets/sec). In fact, the delay aware (ED-DSR$_2$) and the default (ED-DSR$_3$) routing protocol discards the packets that have missed their deadlines in order to reduce the network load with inutile traffic and selects route minimizing energy cost.

## 7 CONCLUSIONS AND FUTURE WORK

In this paper, we propose an energy and delay aware routing protocol ED-DSR and compare its performance with the well known on-demand ad-hoc protocol DSR. The protocols were evaluated through the NS-2.

The main differences between ED-DSR and other on-demand routing protocols is that ED-DSR allows the real-time data flows to be routed from the source to the destination before the expiration delay. It verifies the real-time constraint validity otherwise it discards real-time packets missing their deadlines; thus, it reduces network load and energy consumption of intermediate mobile nodes. Furthermore, the route selection is done according to energy consumption and queue load of intermediate nodes, too. Cost function is defined based on residual energy, queue length, processing and transmission time of intermediate nodes. The ED-DSR selects the route that avoids overloaded and low power intermediate mobile nodes and reduces the delay for each packet.

Simulation results prove the performance of our proposed routing protocol for different traffic loads and network densities. They indicate that ED-DSR prolongs network lifetime and achieves lower energy dissipation per bit of data delivery, higher volume of packets delivered and lower end-to-end delay. In future works, we plan to study the case of unreliable networks in which the nodes can be broken down quickly.

# REFERENCES

[1] Al khwildi A.N., Khan S., Loo K.K., Al raweshidy H.S: Adaptive Link-Weight Routing Protocol using Cross-Layer Communication for MANET. WSEAS Transactions on communications. Issue 11, Volume 6, November 2007.

[2] Crawley, E., Nair, R., Rajagopalan, B. and Sandick, H: A Framework for QoS-based Routing in the Internet. IETF RFC2386, 1998.

[3] Das S., Castañeda R., Yan J.: Simulation-Based Performance Evaluation of Routing Protocols for Mobile Ad-hoc Networks. Mobile Networks and Applications, 5(3): pp. 179-189. (2000).

[4] Hu Y. C., Johnson D. B.: Implicit Source Routes for On-Demand Ad Hoc Network Routing. ACM MobiHoc, (2001).

[5] David B. J., David A. Maltz, Broch J.: DSR: The Dynamic Source Routing Protocol for Multi-Hop Wireless Ad-hoc Networks. in Ad-hoc Networking, edited by Charles E. Perkins, Chapter 5, pp. 139-172, Addison-Wesley. (2001).

[6] Frikha M., MaamerM.: Implementation and simulation of OLSR protocol with QoS in Ad-hoc Networks. Proc. of the 2nd International Symposium on Communications, Control and Signal Processing. (2006).

[7] Huang J., Chen M., Peng W. : Exploring Group Mobility for Replica Data Allocation in a Mobile Environment. Proc. Twelfth international conference on Information and knowledge management, pp. 161-168. (2003).

[8] Kuo C., Pang A., Chan S.: Dynamic Routing with Security Considerations. IEEE Transactions on Parallel and Distributed Systems, vol. 20, no. 1, pp. 48-58. (2009).

[9] Laouiti A., Adjih C.: Mesures des performances du protocole OLSR, Hipercom. In international conference in Sciences of Electronic, Technology of Information and Telecommunications (SETIT). (2003).

[10] Maleki M., Pedram M.: Power-Aware On-Demand Routing Protocols for Mobile Ad-hoc Networks. In Low Power Electronics Design, Edited by C. Piguet. The CRC Press, (2004).

[11] Mbarushimana C., Shahrabi A.: Congestion Avoidance Routing Protocol for QoS-Aware MANETs. Proc. of International Wireless Communications and Mobile Computing Conference, pp. 129-134; (2008).

[12] Meng L., Lin Z., Victor L., Xiuming S.: An Energy-Aware Multipath Routing Protocol for Mobile Ad Hoc Networks. Proc of Sigcomm Asia Workshop, Beijing, China, pp.166-174, (2005).

[13] Ouni S., Bokri J.,Kamoun F.: DSR based Routing Algorithm with Delay Guarantee for Ad-hoc Networks. JNW 4(5): 359-369, (2009).

# AUTHORS

**Jihen DRIRA REKIK** is a Ph. D. student working on QoS guarantee of real-time database system over mobile Ad-hoc Networks at the National School of Computer Sciences of Tunis. She received her Master in Communication systems from National Engineering School of Tunis (ENIT) and her engineering degree in computer network from National Institute of Applied Science and Technology (INSAT). Now, as a Ph. D. candidate, her research interests include energy efficiency, delay guarantee, mobile ad-hoc networks and real-time database system.

**Leïla BACCOUCHE** received her Ph. D. in computer science from the National Polytechnic Institute of Grenoble in France in 1996. She is an assistant professor at the National Institute of Applied Science and Technology in Tunisia in the computer science and mathematics department. Her research interest is related to real-time systems and real-time databases and includes scheduling, quality of service, feedback control and mobile wireless and ad-hoc networks.

**Henda BEN GHEZALA** is currently a Professor of Computer Science in the department of Informatics at the National School of Computer Sciences of Tunis. She leads a Master degree in 'ICIS'. She is the president of University of Manouba. Her research interests lie in the areas of information modeling, databases, temporal data modeling, object-oriented analysis and design, requirements engineering and specially change engineering, method engineering. She is the director of the RIADI laboratory.